# All-optical cryogenic thermometry based on NV centers in nanodiamonds

M. Fukami[1,†], C. G. Yale[1,†], P. Andrich[1,‡], X. Liu[1], F. J. Heremans[1,2], P. F. Nealey[1,2], D. D. Awschalom[1,2,*]

1. Institute for Molecular Engineering, University of Chicago, Chicago, IL 60637
2. Institute for Molecular Engineering and Materials Science Division, Argonne National Lab, Argonne, IL 60439

[†]Present address: Sandia National Laboratories, Albuquerque, NM, 87185
[‡]Present address: University of Cambridge, Cavendish Laboratory, JJ Thomson Ave, Cambridge CB3 0HE
[*]Email: awsch@uchicago.edu

## ABSTRACT

The nitrogen-vacancy (NV) center in diamond has been recognized as a high-sensitivity nanometer-scale metrology platform. Thermometry has been a recent focus, with attention largely confined to room temperature applications. Temperature sensing at low temperatures, however, remains challenging as the sensitivity decreases for many commonly used techniques, which rely on a temperature dependent frequency shift of NV center's spin resonance and its control with microwaves. Here we use an alternative approach that does not require microwaves, ratiometric all-optical thermometry, and demonstrate that it may be utilized to liquid nitrogen temperatures without deterioration of the sensitivity. The use of an array of nanodiamonds embedded within a portable polydimethylsiloxane (PDMS) sheet provides a versatile temperature sensing platform that can probe a wide variety of systems without the configurational restrictions needed for applying microwaves. With this device, we observe a temperature gradient over tens of microns in a ferromagnetic-insulator substrate (yttrium iron garnet, YIG) under local heating by a resistive heater. This thermometry technique provides a cryogenically compatible, microwave-free, minimally invasive approach capable of probing local temperatures with few restrictions on the substrate materials.

## I. INTRODUCTION

Local temperature variation plays a central role in many-body physics governed by hydrodynamic descriptions [1,2], in biomolecular science [3], as well as in thermal engineering of integrated circuits. Among the existing high-sensitivity nanometer-scale thermometers, nitrogen vacancy (NV) centers in nanodiamonds (NDs) have emerged as promising temperature-sensitive fluorescent probes. The negatively-charged NV-center (NV[-]) consists of a ground state spin triplet manifold with a zero-field splitting $\mathcal{D} \simeq 2.87$ GHz that sensitively responds to temperatures, where the shift can be measured by reading out the spin optically [3–6]. By virtue of diamond's high thermal conductivity and NV[-] centers' long spin coherence time, ND-based thermometry has been demonstrated in a variety of systems, such as within a living cell at room temperature [3]. The temperature response of $\mathcal{D}$ is significantly smaller at low temperatures, however, which reduces sensitivity and hinders the conventional thermometry technique [7,8].

Ratiometric all-optical thermometry has been proposed as an alternative to the conventional microwave spin-resonance thermometry technique with compatible sensitivity at room temperature [9–12]. It also enables temperature sensing without the application of microwaves, which removes concerns of microwave heating. Interestingly, the temperature sensitivity of the all-optical thermometer is estimated to improve at lower temperatures (see Supplementary Material, Sec. A [13]), and indicates that this technique can offer a path forward towards ND-based cryogenic thermometry. The use of an array of NDs on a polydimethylsiloxane (PDMS) sheet [13] combined with all-optical thermometry completely removes configurational restrictions needed for microwave applications, offering a versatile device capable of probing a wide variety of solid-state systems over tens of microns with an adjustable spatial resolution on the order of a few microns. This makes all-optical thermometry suitable for probing and imaging a variety of condensed matter systems, and may have advantages over conventional NV-center thermometry techniques depending on the required thermal or spatial resolutions as well as the potential microwave response of the target system.

Here we extend the all-optical thermometry technique based on the NV[-] centers in NDs from room temperature to liquid nitrogen temperatures, 85 K ≤ T ≤ 300 K, and demonstrate its application on a ferromagnetic insulator (yttrium iron garnet, YIG) substrate. In particular, we focus on YIG as a platform to demonstrate our sensing approach both because the microwaves used to manipulate NV centers in conventional thermometry would impact the magnetic spins in the YIG [14–19], and the low temperature thermal response of YIG is of interest in the study of the spin-Seebeck effect [20–24]. We initially demonstrate that a laser-pulse sequence to control the NV centers' charge states improves the sensitivity of the all-optical thermometer by approximately a factor of $\sqrt{3}$. Next, we systematically study the temperature dependence of the sensitivity, demonstrating that it improves at cryogenic temperatures. Finally, we apply this all-optical cryogenic thermometry technique at $T = 170$ K to measure the surface temperature profile of a YIG slab in contact with a resistive heater, with the array of NDs embedded on the surface of a flexible PDMS sheet. The observed temperature gradient over a range of tens of micrometers confirms the applicability of the technique on the YIG substrate,indicating that it provides a tool for studying local thermal properties of a wide variety of substrates over a broad range of temperatures.

## II. DEMONSTRATION OF CRYOGENIC ALL-OPTICAL THERMOMETRY

We focus on the temperature dependence of the $NV^-$ centers' zero phonon line (ZPL) amplitude ratio ($A$), which is defined as the ratio of the ZPL intensity with respect to an average photoluminescence (PL) intensity in a spectral range around the ZPL. The ratio $A$ strongly responds to temperature change due to the presence of a coupling between the orbital state of $NV^-$ and vibrational modes in diamond [25] (see Supplementary Material, Sec. B [13]), which leads to a high temperature sensitivity. The experiment was conducted on an array of NDs containing ensembles of $NV^-$ centers measured with a confocal microscope using a high numerical aperture objective (NA=0.9) as shown in Fig. 1(a). An array of NDs embedded into the flexible PDMS sheet was placed on the surface of a 3.05-μm-thick YIG film grown on a 500-μm-thick gadolinium gallium garnet (GGG) substrate (MTI Corp.). A Ti/Au (thickness: 8nm/200nm) resistive heater, for local heating, was patterned on the YIG film using a lithographic process. The bottom of the GGG substrate was affixed to a copper thermal sink within a flow cryostat. Both characterization (section II) and application (section III) of the thermometry were conducted on the same device with a YIG substrate for consistency (for data without a PDMS sheet on a quartz substrate, see Supplementary Material, Sec. I [13]).

Figure 1(b) shows a two-dimensional PL scan of an individual spot in the array of NDs under continuous 594-nm excitation measured by an avalanche photodiode (APD). The 594-nm light does not excite the neutrally-charged NV-center ($NV^0$) [26,27] and removes the noisy $NV^0$ phonon-sideband spectral emission from the $NV^-$'s ZPL spectrum. The diameter of the spot is 1000 nm which is defined by our microfabrication technique [28], and contains tens of NDs, where each ND contains hundreds of NV centers [28]. Figure 1(c) shows a horizontal cut through the maximum of Fig. 1(b). Interestingly, when we applied pulse sequences of the 594-nm and 532-nm lasers as shown in Fig. 1(d), which is in contrast to the previous studies with a continuous-wave excitation [11,12], the PL count rate was enhanced by approximately a factor of three (see Supplementary Material, Sec. D [13]). The enhancement is due to the charge-state conversion between $NV^-$ and $NV^0$ [13,29–31]. While charge-state conversions of NV centers in NDs have not been comprehensively studied to our knowledge, we simply assume the results reported in bulk diamonds are applicable and attribute the PL enhancement to the charge-state conversion. Since the sensitivity of the all-optical thermometer is limited by shot noise, improving the PL count rate by a factor of $\approx 3$ increases sensitivity by a factor of $\approx \sqrt{3}$ (see Supplementary Material, Sec. E [13]). In the following spectral measurements, we send the PL to a spectrometer and gate the intensifier of a single-photon sensitive CCD camera in the spectrometer (iStar 334T, Andor) triggered by the pulse sequences. Every spectral measurement was followed by a background measurement taken off the ND and the background counts were subtracted. (see Supplemental Material, Sec. F [13]).

Figure 2(a) shows the PL spectra $L(h\nu)$ of $NV^-$ centers in the temperature range 85 K $\leq T \leq$ 100 K. Monotonic change in the spectra is observed except near $T \simeq 230$ K and $T \simeq 150$ K, which are due to the melting point and the glass transition point of the PDMS, respectively. We note that the presence of the PDMS sheet does not change the thermometry property of NV centers except PL count rates, which is verified by the measurements done on NDs without a PDMS sheet (See Supplementary Material, Sec. I [13]). To maximize the PL count rate, we widely opened the slit in the spectrometer, which results in a wavelength resolution $\delta\lambda = 3.5$ nm. For the temperature sensing, we focus on the ZPL emission peak at $h\nu \simeq 1.94$ eV (637 nm). Importantly, the ZPL becomes sharper and more prominent at lower temperatures. In this experiment, we focused on the PL in the wavelength ranging from 605 nm to 660 nm, which we define as the spectral range ($\mathcal{R}$) (for the choice of this range, see Supplementary

Material, Sec. G [13]). As shown in the inset of Fig. 2(b), we fit the relative spectrum $L/\langle L \rangle_\mathcal{R}$ by a sum of a squared-Lorentzian function and an exponential function

$$L(h\nu)/\langle L \rangle_\mathcal{R} = A \frac{1}{[w^2 + (h\nu - h\nu_{ZPL})^2]^2} + B \exp\left[-\frac{h(\nu - \nu_{ZPL})}{k_B \Theta}\right] \quad (1)$$

where $k_B$ is the Boltzmann constant, $h$ is the Plank constant, $\langle L \rangle_\mathcal{R}$ is the average PL intensity in the spectral window $\mathcal{R}$ and $\{A, B, \Theta, w, \nu_{ZPL}\}$ are fitting parameters. A squared-Lorentzian function instead of a Lorentzian function is used as suggested in Ref. [32] for better fits at cryogenic temperatures. Temperature dependence of the ratio $A$ is shown as solid markers in Fig. 2(b), where the solid and dotted curves are derived from the two fits of the reduced Debye-Waller factor and the ZPL linewidth shown in Figs. 2(c) and 2(d). We note that the reduced Debye-Waller factor is defined in this work as the ratio of the integrated ZPL emission, which corresponds to the area under the squared-Lorentzian fit, to the total PL in the range $\mathcal{R}$. Importantly, we find a maximum in the slope of the ratio $A$ around $T \sim 150$ K, which coincidentally corresponds to the glass transition temperature of the PDMS, though does not appear to be related to it (See Supplementary Material, Sec. H and Sec. I [13]).

While the stronger temperature response $|dA/dT|$ at lower temperatures observed in this study is desirable for the improved temperature sensing, the presence of the maximum cannot be explained by a currently existing model, since it predicts a monotonic increase of the temperature response at lower temperatures. This can be resolved by taking into account a constant term ($a$) in the linewidth $w = a + bT^2$, modifying the analytical expression of the ZPL amplitude ratio to be (see Supplemental Material, Sec. J [13])

$$A = \frac{2\alpha \exp(-\gamma T^2) \Delta \mathrm{R}}{\pi(a + bT^2)} \quad (2)$$

where $\alpha$ and $\gamma$ are fitting parameters of the reduced Debye-Waller factor and $\Delta \mathrm{R}$ is the size of the spectral window $\mathcal{R}$. The constant contribution is due both to a resolution $\delta\lambda$ of the spectrometer and an inhomogeneous broadening. Wavelength resolution $\delta\lambda$ can be improved by narrowing down the slit in the spectrometer with a trade-off of the PL count rate. The inhomogeneous broadening is not negligible at lower temperatures due to crystal strain variations both between different NDs and within the individual commercial NDs used in this study. These limitations could be overcome by introducing engineered nanoparticles [33,34], leading to an enhanced temperature response at cryogenic temperatures.

The temperature sensitivity $\eta$ of a thermometer, which is sometimes referred to as the noise floor, is not only quantified by the temperature response $|dA/dT|$ but also by the uncertainty $\sigma_A$ in the measurement of $A$. They are related by $\eta = \sigma_A \sqrt{\Delta t} |dA/dT|^{-1}$, where $\Delta t$ is the measurement time. While the temperature response increases at lower temperatures, $\sigma_A$ grows along with the temperature response. To fully characterize the sensitivity of the thermometry technique, we studied the uncertainty $\sigma_A$ as a function of temperature $T$. At each temperature, PL spectrum measurements with an integration time of $\Delta t = 2.5$ s were repeated one hundred times (Fig. 3(a)). We then calculate the standard deviation $\sigma_A$ for each data set and show its temperature dependence in Fig. 3(b). Note that the standard deviation $\sigma_A$ is rescaled by a factor $\sqrt{C_{ZPL} \Delta t}$ to quantitatively compare the results at different temperatures, where $C_{ZPL}$ is the ZPL count rate shown in the inset of Fig. 3(c) that corresponds to the area under the squared-Lorentzian fit (see Supplemental Material, Sec. L [13]). The dashed curve shows the lower bound when the noise is coming only from photon shot noise, while the dotted curve shows the lower bound when the CCD camera's dark-current shot noise also contributes to the noise in the measurement of the ratio $A$ (see Supplementary Material, Sec. M [13]). The experimental observation is well explained by the dotted curve, demonstrating that the standard deviation $\sigma_A$ is limited both by the ZPL photon shot noise and the CCD's dark current shot noise.

Combining the temperature dependencies of $\sigma_A$ and $|dA/dT|$ as shown in Figs. 3(c) and 2(b), we plot the temperature dependence of the sensitivity $\eta$ in Fig. 3(d). The lower bounds shown are derived from the same models as in Fig. 3(c). Importantly, the sensitivity improves at cryogenic temperatures in contrast to the conventional thermometry technique based on the temperature dependent shift in the zero-field splitting. We note that the sensitivity calculated in this study at $T = 300$ K does not reach the level of the sensitivity provided in the previous report on all-optical thermometry at room temperature [11]; however, taking into account detection efficiency differences, our result is found to be fully consistent with the one in Ref. [11]. This can be confirmed by introducing a projected sensitivity $\eta_{proj}$ as shown in Fig. 3(d), which assumes as high ZPL counts rates as in Ref. [11] and shows an anticipated sensitivity compatible with their result (for detail, see Supplemental Material, Sec. O [13]). The highest temperature sensitivity is achieved near $T \sim 200$ K, which can be understood through the simplified analytical model that only considers the temperature evolution of the DWF (for detail on the necessary assumptions, see Supplemental Material, Sec. P [13])

$$\eta \sim \frac{1}{\gamma\sqrt{2C_{tot}\,(\text{DWF})\big|_{T=0}}} T^{-1} \exp\left(\frac{1}{2}\gamma T^2\right) \quad (3)$$

resulting in a minimum at $T \sim 1/\sqrt{\gamma} = 218$ K, where $C_{tot}$ is the total PL counts rate of NV$^-$ and $(\text{DWF})\big|_{T=0}$ is the (non-reduced) Debye-Waller factor at absolute zero (For the discussion of the effect of the PDMS sheet, see Supplemental Material, Sec. Q [13]). While there is a quantitative mismatch due to oversimplification in the model, this model captures the existence of the minimum well. To further improve the sensitivity at low temperatures, one could, for instance, increase the ZPL count rate by improving the detection efficiency and utilize brighter NDs that contain more NV$^-$ centers.

### III. SURFACE TEMPERATURE IMAGING OF A YIG FILM

To demonstrate the applicability of the all-optical thermometer, we apply an 80-mA current to the resistive heater to generate a temperature gradient in the YIG and measure the spatial temperature variation of the YIG surface using an array of NDs, as illustrated in Fig. 1(a). Since the YIG has spin-wave resonances at microwave frequencies near $\mathcal{D}$ [14–19], this measurement confirms that the all-optical thermometry technique can be used independently of substrate materials where microwave control is problematic. In these experiments, the base temperature of the copper heat sink is stabilized at $T = 170$ K (see Supplemental Material, Sec. R [13]). Figure 4(a) shows a two-dimensional spatial scan of the PL from the array of NDs used in this study. To construct the temperature profile, we repeat temperature measurements at multiple spots in the array. The accuracy of the measured temperature is ensured by calibrating NDs individually (see Supplemental Material, Sec. S [13]) and the temperature dependencies of $\{B, \Theta, w, \nu_{ZPL}\}$ in addition to $A$ are utilized for calculating the local temperature (see Supplemental Material, Sec. T [13]). For each measurement, the PL is collected in total for 500 s.

Figure 4(b) shows the resulting temperature profile of the YIG surface, where we observe a temperature decay on the order of tens-of-microns from the heat source. The temperature of each spot as a function of the distance from the heater is shown in Fig. 4(c), where the error bars include both the uncertainty of the sensing and the error in the calibration. The data is fit well by the Green's function to the two-dimensional Poisson equation, showing that the temperature field in the YIG approximately follows the steady state diffusion equation with a single heat carrier. We note that the Poisson equation is not accurate in YIG because there are two kinds of heat carriers, phonons and magnons. A deviation from the Poisson equation is expected near the heat source within a length scale of a magnon-phonon thermalization, which is much smaller than a few micrometers [35]. In our experiment, NDs directly measure temperatures of the YIG lattice, or phonons, and we do not observe any perturbation to the qualitative feature of the steady-state phononic temperature profile by the presence of magnons in YIG, which is expected due to our thermal and spatial resolutions. (see Supplemental Material, Sec. U [13]) [20,24].

### IV. CONCLUSION

We demonstrate and characterize an all-optical thermometry technique based on NV$^-$ center ensembles in ND that can be deployed from room temperature to liquid nitrogen temperatures, with a sensitivity that increases with decreasing temperature. Furthermore, the PL intensity of NV$^-$ centers is enhanced by implementing pulse sequences to convert NV$^0$ into NV$^-$, leading to a higher temperature sensitivity by approximately a factor of $\sqrt{3}$. Systematic noise analysis reveals that the sensitivity is limited by the shot noise and the inhomogeneous broadening of the ZPL linewidth, suggesting a pathway for further sensitivity improvements by optimizing the spectral resolution, improving the PL detection efficiency, and introducing engineered NDs with high brightness and homogeneous crystal strains. Taking advantage of an array of NDs embedded in a flexible PDMS sheet, we show the utility of the all-optical thermometer at $T$ =170 K by measuring the surface temperature profile of a YIG slab thermally driven by a resistive heater. This all-optical thermometry technique along with the versatility of the ND membrane array provides a microwave-free, minimally invasive, and cryogenically compatible way of measuring local temperatures within a variety of substrate materials.

**ACKNOWLEDGMENTS**

This work was supported by the Air Force Office of Scientific Research and the Army Research Office through the MURI program, grant no. W911NF-14-1-0016. The fabrication of the diamond nanoparticle arrays was supported by the US Department of Energy, Office of Science, Basic Energy Sciences, Materials Sciences and Engineering Division. FJH, PFN, and DDA were supported by the US Department of Energy, Office of Science, Basic Energy Sciences, Materials Sciences and Engineering Division. This work made use of shared facilities supported by the NSF MRSEC Program under grant no. DMR-0820054. The authors thank P. C. Jerger, B. B. Zhou, C. M. Anderson and J. C. Karsch for useful discussions.

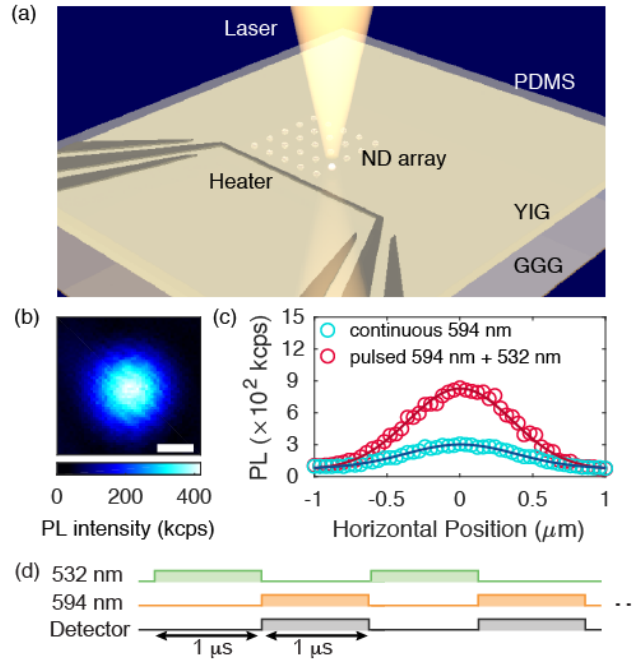

FIG. 1. (a) Schematic of an array of nanodiamonds (NDs) on a 3.05-µm YIG film grown on a GGG substrate. NDs are embedded on the surface of a flexible PDMS sheet and the YIG film was patterned with a resistive heater (central wire has a width of 5 µm and a length of 200 µm). (b) Two-dimensional photoluminescence (PL) image of NV centers in NDs collected under continuous 594-nm excitation. PL intensity is measured by an avalanche photodiode (APD). The measurement was conducted at $T=170$ K. Scale bar, 0.5 µm. (c) Line cuts of PL intensity profiles of NV centers under two different excitation pulse sequences. (d) Schematic of the pulse sequences of a 532-nm laser (NV⁻ charge state initialization), a 594-nm laser (NV⁻ detection) and a detector (APD/CCD camera).

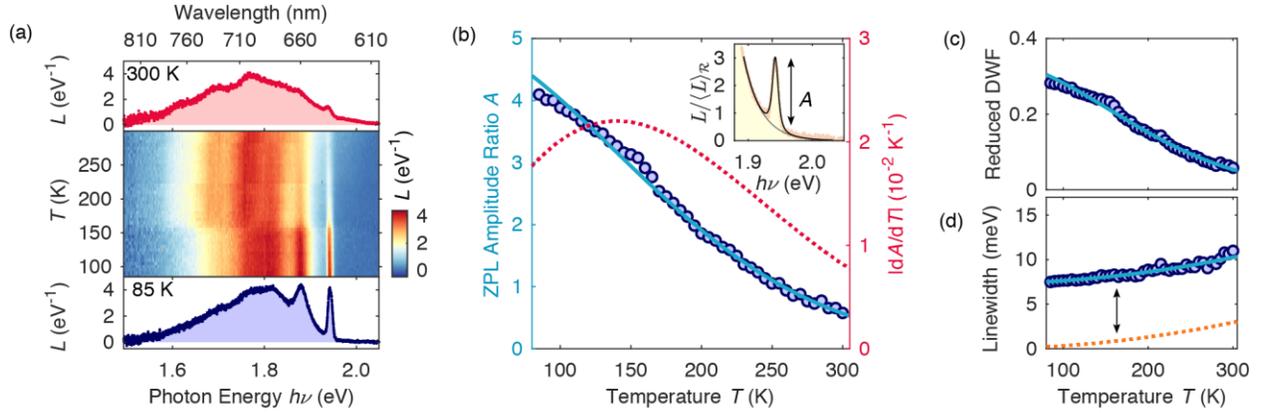

FIG. 2. (a) Evolution of NV centers' PL spectrum $L(h\nu)$ between temperatures $T$ =85 K and $T$ =300 K. The areas under the spectra are normalized to one. Discontinuities at $T \simeq 230$ K and $T \simeq 150$ K are associated with the PDMS's phase transitions and not related to NV centers. Top (bottom) graph shows the spectrum at 300 K (85 K). (b) Temperature dependence of the ZPL amplitude ratio $A$ (left axis) and its temperature response $|dA/dT|$ (right axis). The solid blue curve is calculated from two fits: (i) temperature dependence of the reduced Debye-Waller factor (DWF) (shown in (c)), and (ii) temperature dependence of the ZPL linewidth (shown in (d)). The dotted red curve is the derivative of the solid (blue) curve with respect to temperature $T$. Inset shows the fit of the ZPL at $T$ =170 K with a sum of an exponential function and a squared-Lorentzian function (black curve). The exponential-function part only is shown with a gray curve. $\langle L \rangle_\mathcal{R}$ is the mean PL intensity in the range $\mathcal{R}$ from 605 nm to 660 nm. (c) Reduced DWF as a function of temperature $T$. A Gaussian-functional fit is shown. (d) ZPL linewidth as a function of temperature $T$. The solid blue fit is the second-order polynomial $a+bT^2$ and the dotted orange curve shows $bT^2$.

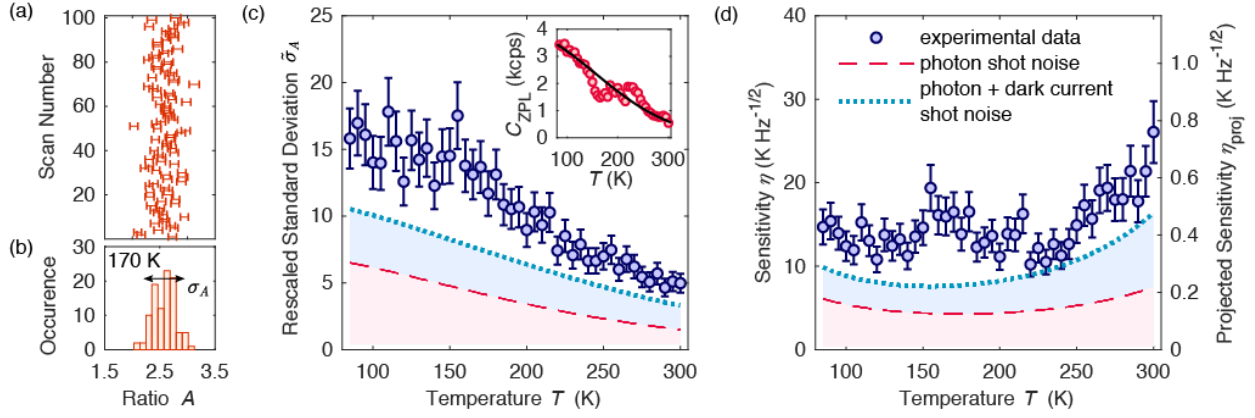

FIG. 3. (a) ZPL amplitude ratio scanned over 100 times. A single scan consists of a total PL accumulation time of 2.5 s. The measurements were conducted at temperature $T = 170$ K. (b) Histogram built from the measurements in (a). The standard deviation $\sigma_A$ is depicted. (c) Rescaled standard deviation $\tilde{\sigma}_A = \sigma_A \sqrt{C_{ZPL} \Delta t}$ as a function of temperature $T$, where $C_{ZPL}$ is the PL counts rate under the squared-Lorentzian fit of ZPL and $\Delta t$ is the total PL accumulation time. The dashed red curve shows the lower bound determined by photon shot noise and the dotted blue curve shows the lower bound determined by photon and dark current shot noise. Inset shows ZPL counts rate $C_{ZPL}$ as a function of temperature $T$. Solid black curve shows a one-parameter ($a_1$) fit of the ZPL counts rate $C_{ZPL}(T) = a_1 (DWF)_R^{(T)}$, where $(DWF)_R^{(T)}$ is the curve shown in Fig. 2 (c). (d) Temperature sensitivity $\eta$ as a function of temperature $T$. The dashed red and the dotted blue curves identify the lower bounds for the sensitivity as defined in (c). The spike near 160 K arises from the dip in the experimental data of $C_{ZPL}$ as shown in the inset of (c). Right axis shows a projected sensitivity $\eta_{proj}$ under the assumption of a higher detection rate of the PL as explained in the main text.

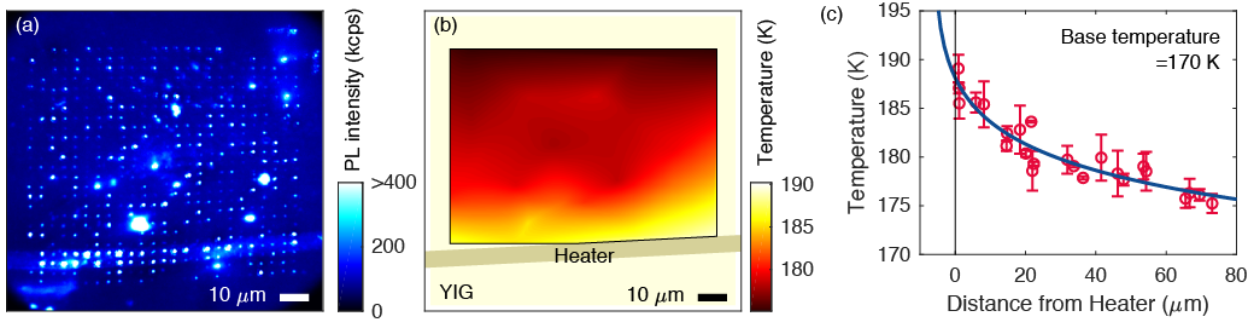

FIG. 4. (a) Spatial PL scan of NV centers in NDs in the array. (b) Two-dimensional temperature imaging of the YIG surface using NV centers in the array of NDs embedded on the surface of the PDMS sheet measured by the all-optical thermometry technique. An 80-mA current is applied to the resistive heater. The base temperature was set to $T = 170$ K. (c) YIG surface temperature as a function of the distance from the resistive heater. Fit with a logarithmic function is shown.

# Supplemental Information for
# All-optical cryogenic thermometry based on NV centers in nanodiamonds

M. Fukami[1], C. G. Yale[1,†], P. Andrich[1,‡], X. Liu[1], F. J. Heremans[1,2], P. F. Nealey[1,2], D. D. Awschalom[1,2,*]

1. Institute for Molecular Engineering, University of Chicago, Chicago, IL 60637
2. Institute for Molecular Engineering and Materials Science Division, Argonne National Lab, Argonne, IL 60439

†Present address: Sandia National Laboratories, Albuquerque, NM, 87185
‡Present address: University of Cambridge, Cavendish Laboratory, JJ Thomson Ave, Cambridge CB3 0HE
*Email: awsch@uchicago.edu

### A. Temperature Dependence of the Sensitivity in a Range $300 \text{ K} \lesssim T \lesssim 400 \text{ K}$

Reference [S1] provides a model that explains the temperature dependence of the zero-phonon line (ZPL) amplitude ratio $A$ under temperature $T$ in the range $300 \text{ K} \lesssim T \lesssim 400 \text{ K}$. The authors fit the ZPL with a sum of a Lorentzian function and an exponential function, with the coefficients, $A$ and $B$, respectively. In the model, the ratio $A$ is proportional to the Debye-Waller factor (DWF) divided by a ZPL linewidth $w$. Then the temperature dependence of the ratio $A$ is given by

$$A = \alpha T^{-2} \exp(-\gamma T^2), \qquad (S1)$$

resulting in the temperature response

$$\left|\frac{dA}{dT}\right| = 2T(T^{-2} + \gamma)A, \qquad (S2)$$

where $\alpha$ and $\gamma$ are temperature independent constants which are related to the electron-phonon coupling $S$, the Debye temperature $T_D$ and reference values. We note that the DWF is defined as the ratio of the integral ZPL intensity to the total PL. From this expression, the temperature response is expected to be larger at lower temperatures, which potentially gives rise to a higher temperature sensitivity at lower temperatures though it also depends on the uncertainty of the measurement of the ratio $A$. The uncertainty $\sigma_A$ is given by [S2]

$$\sigma_A = \frac{f(r)A}{\sqrt{C_{\text{ZPL}}\Delta t}} \qquad (S3)$$

$$f(r) = \sqrt{c_1 + c_2 r + c_3 \sqrt{r^2 + r}}, \qquad \text{with } [c_1, c_2, c_3] = [3,3,1], \qquad (S4)$$

where $r = B/A$, $C_{\text{ZPL}}$ is the ZPL counts rate and $\Delta t$ is the measurement time. From the equation (S3), the temperature sensitivity, or the noise floor, can be written as

$$\eta \equiv \sigma_A \sqrt{\Delta t} |dT/dA| = \frac{Tf(r)}{2(1+\gamma T^2)\sqrt{C_{\text{ZPL}}}}. \qquad (S5)$$

Assuming, for simplicity, that the temperature dependence of the total PL is negligible, we can write

$$C_{ZPL} = C_{tot}(\text{DWF})|_{T=0} \exp(-\gamma T^2), \qquad (S6)$$

where $C_{\text{tot}}$ is the total PL counts rate and $(\text{DWF})|_{T=0}$ is the DWF at absolute zero. From equations (S5) and (S6), we get

$$\eta = \frac{Tf(r)}{2(1+\gamma T^2)\sqrt{C_{\text{tot}}(\text{DWF})|_{T=0}}} \exp\left(\frac{1}{2}\gamma T^2\right). \qquad (S7)$$

As the temperature decreases, the factor $r = B/A$ decreases, which is not shown in the Ref. [S1] but is confirmed in a regime $85 \text{ K} \leq T \leq 300 \text{ K}$ as shown in the Fig. S6(b). Then we get $d\eta/dT > 0$, demonstrating a higher sensitivity at lower temperatures, at least in a regime $300 \text{ K} \lesssim T \lesssim 400 \text{ K}$ where the model is confirmed.

### B. Temperature response of the ratio $A$

The model described in Sec. A assumes that the temperature response of the ratio $A$ is dominated by that of the DWF and the ZPL linewidth. Another possible contribution is the temperature response of the amount of phonon-sideband emission in the range of interest (in our case, the spectral range $\mathcal{R}$) with regard to the total PL, though its temperature dependence is negligible as shown in Sec. I.

### C. Temperature stability of the flow cryostat

In the experiment, the base temperature of the sample was stabilized with PID control. Temperature deviation was within ±0.3 K for all measurements. Though the thermocouple was positioned a few centimeters away from the sample position, temperature accuracy within ±0.5 K was ensured in a calibration of the setup which has a thermocouple right next to the sample position.

### D. Enhancement of the PL at Different Spots in the Array

Figure. 1(c) in the main text shows the enhancement of the PL at $T = 170$ K with the pulse sequences shown in Fig. 1(d) in the main text. In the Fig. S1, we show the enhancement of the PL at different spots in the array. The figure S1(a) is identical to the Figure. 1(c) in the main text, while figures S1(b) and S1(c) show the PL scans at other spots. Each PL peak was fit by a sum of a Gaussian function and a constant, where the amplitudes of the Gaussian functions were extracted from the fits. The enhancement in the amplitudes due to the pulse sequences was observed, where the enhancement factors were approximately 3.1, 2.4, and 3.4 for Figs. S1(a), S1(b) and S1(c), respectively. Though the factor depends on the spots, enhancements by approximately a factor of three were observed.

The enhancement is due to the charge-state conversion between $NV^-$ and $NV^0$. While the 594-nm excitation preferentially converts $NV^-$ into $NV^0$, the 532-nm excitation preferentially converts $NV^0$ back into $NV^-$ [S3–S5]. The time scale of the charge-state conversion depends on the laser power [S4]. To minimize the heating while keeping the PL counts high enough in our experiments, the powers of the two lasers were both set to 200 $\mu$W, leading to an estimate that the relaxation time of the charge-state conversion is larger than 1 $\mu$s.

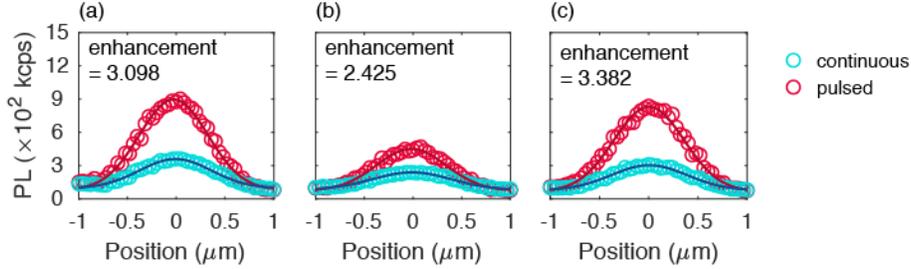

FIG. S1. Enhancement of the PL due to the pulse sequences as shown Fig. 1(d) in the main text under three different spots in the array. (a) shows the same figure as the Fig. 1(c) in the main text.

### E. Discussion of the practical sensitivity under the pulse sequence

Since the sensitivity of the all-optical thermometer is limited by a shot noise, a higher PL count rate by roughly a factor of three results in a higher sensitivity of temperatures by approximately a factor of $\sqrt{3}$. Note that the pulse sequences also reduce the fraction of the measurement time in the total scanning time. While it improves the physical sensitivity $\eta$ of temperatures, it may result in worse practical sensitivity $\eta_\text{practical}$ if the enhancement factor is less than two. We observed, however, improved sensitivity with the pulse sequences not only because the enhancement is larger than two but also because it reduces the noise due to the CCD dark counts and the background counts. Here we note that physical sensitivity $\eta$ is defined as the minimum temperature difference that can be resolved by a given amount of NV-center-measurement time, while the practical sensitivity $\eta_\text{practical}$ is the minimum temperature difference that can be resolved by a given total time including the time necessary for charge state preparation, background measurement, control of the equipment, and feedback control to focus on the target spot.

### F. Background Measurement

Each spectral measurement was followed by an off-spot background measurement with the same measurement duration. This not only deteriorates the practical sensitivity $\eta_\text{practical}$, but also adds additional noise to the physical sensitivity $\eta$. The factor is considered in the calculation of the noise model where CCD camera's dark-current shot noise also contributes in addition to the ZPL photon shot noise, as explained in Sec. L.

### G. Choice of the Range $\mathcal{R} = \{h\nu|\ hc(660 \text{ nm})^{-1} \leq h\nu \leq hc(605 \text{ nm})^{-1}\}$

The spectral range $\mathcal{R}$ is chosen such that it is consistent with previous reports [S1,S6,S7]. With a choice of the grating in our spectrometer (Acton SP-2750, Princeton Instrument, 300 gr/mm with 750 nm blaze; iStar 334T, Andor), the range can be measured in the CCD with a single scan. This allowed us to take measurements without stitching different spectral scans under different angles of the grating in the spectrometer. In contrast, the spectra shown in the Fig. 2(a) in the main text are stitched over multiple scans under different angles of the grating.

### H. Fitting the reduced Debye-Waller factor and the ZPL linewidth

To get a curve for the ratio $A$ in the main text, we fit temperature dependencies of a reduced Debye-Waller factor $(DWF)_\mathcal{R}$ and a ZPL linewidth $w$. We fit the temperature dependence of $(DWF)_\mathcal{R}$ by a Gaussian function $(DWF)_\mathcal{R} = \alpha \exp(-\gamma T^2)$, where $\alpha$ and $\gamma$ are fitting parameters but $\gamma$ is related to the electron-phonon coupling $S$ and the Debye temperature $T_D$ by a relation $\gamma = 2\pi^2 S / 3T_D^2$ [S8]. In our experiment, we measured $\gamma = (218 \text{ K})^{-2}$ which corresponds to $T_D/\sqrt{S} = 560$ K. The value of $\gamma$ was consistent with a measurement conducted on NDs without the PDMS sheet as shown in Sec. I. The temperature dependence of the ZPL linewidth $w$ is fit by a second-order polynomial $w = a + bT^2$. As shown in Figure 2(d) in the main text, the constant contribution $a$ is not negligible at lower temperatures in our experiment due to the inhomogeneous broadening. Based on the two fits, we obtained the curve in Fig. 2(b). Based on the model under a simplifying assumption $a \gg bT^2$, one can easily find that the temperature response $|dA/dT|$ takes maximum at $T \simeq 1/\sqrt{2\gamma} = 154$ K, which is consistent with the experimental observation.

### I. Temperature Dependencies of the Parameters without the PDMS Sheet

The dependency of the ratio $A$ on the temperature shown in Fig. 2(b) in the main text is not largely affected by the presence of the PDMS sheet. To support this statement, we show the PL spectra of NV centers without the PDMS sheet in the Fig. S2 where the spectra were measured at $T = 85$ K, $110$ K, $150$ K, $200$ K, $250$ K and $300$ K. The measurement was conducted on NDs scattered on a quartz substrate, where hundreds of NDs existed under a laser-focused spot. The same analysis as in the main text is conducted. From the fit of the $(DWF)_\mathcal{R}$ as shown in Fig. S3(a) we got $\gamma = (283 \text{ K})^{-2}$, which stays within 25% from the value $\gamma = (218\ K)^{-2}$ of the NDs embedded in the PDMS array, showing that the presence of the PDMS sheet does not change the main result in this report.

When we fit the reduced Debye-Waller factor with a Gaussian function in the main text, there was an implicit assumption that the temperature dependence of $(DWF)_\mathcal{R}$ is approximately that of $(DWF)$ since the temperature dependence of the DWF is known to be

$$(DWF) = \exp\left(-S\left(1 + \frac{2\pi^2 T^2}{3T_D^2}\right)\right). \tag{S8}$$

Figure S3(a) compares $(DWF)$ and $(DWF)_\mathcal{R}$ scanned on NDs without the PDMS sheet. Figure S3(b) shows almost constant ratio $(DWF)/(DWF)_\mathcal{R}$ over the temperature range $85 \text{ K} \leq T \leq 300$ K, confirming the assumption.

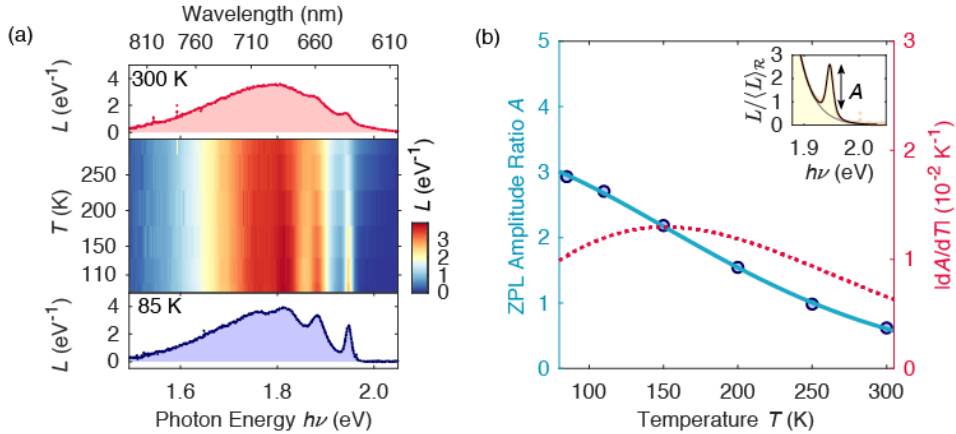

FIG. S2. PL spectra and ZPL amplitude ratio of NV centers without PDMS sheet scanned under multiple temperatures $T = 85$ K, $110$ K, $150$ K, $200$ K, $250$ K and $300$ K. (b) inset shows the spectrum at $T = 150$ K.

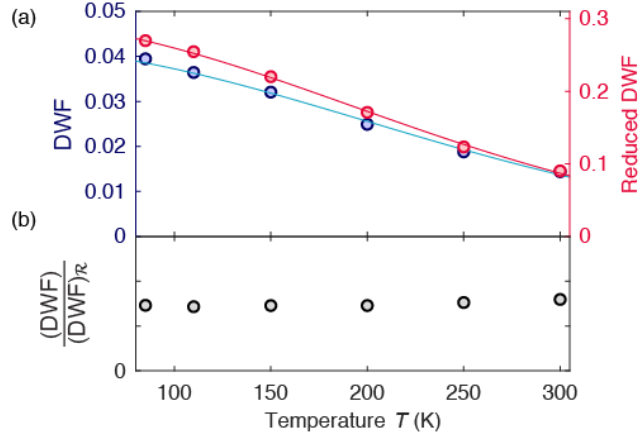

FIG. S3. (a) Temperature dependencies of the Debye-Waller factor (DWF) (left axis) and the reduced Debye-Waller factor $(\mathrm{DWF})_\mathcal{R}$ (right axis) measured on NDs without the PDMS sheet. (b) Temperature dependence of the ratio $(\mathrm{DWF})/(\mathrm{DWF})_\mathcal{R}$, which equals to the fraction of the integrated PL in the range $\mathcal{R}$ to the total PL.

### J. Temperature Dependence of the ZPL Amplitude Ratio

In the fit of the spectrum as shown in the inset of the Fig. 2(b) in the main text, the PL intensity $L(h\nu)$ was firstly divided by the mean PL intensity in the range $\mathcal{R} = \{h\nu|\ hc(660\ \mathrm{nm})^{-1} \leq h\nu \leq hc(605\ \mathrm{nm})^{-1}\ \}$. The mean PL intensity $\langle L \rangle_\mathcal{R}$ can be explicitly written as

$$\langle L \rangle_\mathcal{R} = \frac{1}{\Delta \mathcal{R}} \int_{\nu_\mathrm{i}}^{\nu_\mathrm{f}} L(h\nu) h d\nu, \tag{S9}$$

where $\nu_\mathrm{i} = c(660\ \mathrm{nm})^{-1}$, $\nu_\mathrm{f} = c(605\ \mathrm{nm})^{-1}$ and $\Delta \mathcal{R} = h(\nu_\mathrm{f} - \nu_\mathrm{i})$. Therefore, the reduced Debye-Waller factor $(\mathrm{DWF})_\mathcal{R}$, the ZPL amplitude ratio $A$, and the linewidth $w$ are related by

$$(\mathrm{DWF})_\mathcal{R} = \frac{\frac{\pi}{2} A w}{\Delta \mathcal{R}}. \tag{S10}$$

We note that the reduced Debye-Waller factor is defined as the ratio of the integrated ZPL emission intensity to the total PL in the range $\mathcal{R}$. With the use of the coefficients from the fits of $(\mathrm{DWF})_\mathcal{R}$ and $w$, the ZPL amplitude ratio $A$ can be written as

$$A = \frac{2\alpha \exp(-\gamma T^2)\, \Delta \mathcal{R}}{\pi(a + bT^2)}. \tag{S11}$$

From the equation (S11), we get

$$\left|\frac{dA}{dT}\right| = 2T\left(\frac{b}{a+bT^2} + \gamma\right) A. \tag{S12}$$

### K. Discussion of the value $T_\mathrm{D}/\sqrt{S}$

From the fit of the reduced Debye-Waller factor in the main text, we obtained $T_D/\sqrt{S} = 560$ K. Though this is relatively small considering the bulk Debye temperature $T_\mathrm{D}^\mathrm{bulk} \simeq 2200$ K of diamond, Debye temperatures in nanodiamonds are known to be around 30% smaller [S6]. Mismatch from the literature value of nanodiamonds $T_\mathrm{D}/\sqrt{S}\big|_\mathrm{literature} = 1.0(1) \times 10^3$ K given in Ref. [S1] would be due to the different ensemble of NVs used in our experiment. While tens of commercial NDs with 100-nm diameter were used in this study, Ref. [S1] reports measurements on a single ND with diameter smaller than 50 nm prepared from synthetic sub-micron diamond powder. In the measurement without the PDMS sheet shown in Fig. S3(a), we got similar value $\gamma = (283\ \mathrm{K})^{-2}$ which corresponds to $T_\mathrm{D}/\sqrt{S} = 725$ K. This supports that the smaller value of $T_\mathrm{D}/\sqrt{S}$ measured in our experiment compared to the value in Ref. [S1] is due to the different ensembles of NVs.

### L. Temperature Dependence of the ZPL Counts Rate

There is a subtlety in modeling the temperature dependence of the ZPL counts rate $C_\mathrm{ZPL}$ because it is not only determined by the temperature dependence of the NV center's optical lifetime, but also affected by the temperature dependencies of the steady state population of $\mathrm{NV}^-$ and the PDMS sheet's optical transparency in our experiment. For simplicity, we conducted a one-parameter ($a_1$) fit of the ZPL counts rate $C_\mathrm{ZPL}(T) = a_1 (\mathrm{DWF})_\mathcal{R}^{(T)}$ where $(\text{DWF})_{\mathcal{R}}^{(T)}$ is the curve we got in Fig. 2(c) in the main text. The underlying assumption is that the temperature dependence of $C_{\text{ZPL}}$ is dominated by that of the reduced DWF, which is valid when the integrated PL intensity in the range $\mathcal{R}$ is not significantly temperature dependent compared to the temperature dependence of $(\text{DWF})_{\mathcal{R}}$. A detailed study of the temperature dependence of the ZPL counts rate is beyond the scope of this report.

### M. Two Models for the Rescaled Standard Deviation

The dotted and dashed curves in the Fig. 3(c) in the main text show lower bounds for the rescaled standard deviation under different models. The dashed curve is the lower bound when the noise is coming only from the photon shot noise of the ZPL and the phonon sideband under the ZPL, while the dotted curve shows the lower bound when the CCD camera's dark-current shot noise also contributes to the uncertainty $\sigma_A$. In a model where photon counts under the ZPL peak add noise to the fit of the ZPL, the rescaled standard deviation $\sigma_A\sqrt{C_{\text{ZPL}}\Delta t}$ is given by

$$\sigma_A\sqrt{C_{\text{ZPL}}\Delta t} = g(y)A \quad (S13)$$

$$g(y) = \sqrt{c_1 + c_2 y + c_3\sqrt{y^2 + y}}, \quad \text{with } [c_1, c_2, c_3] \simeq [2.00, 1.98, 0.763]. \quad (S14)$$

The function $g(y)$ differs from the case when using a Lorentzian function [S2]. The dashed curve is drawn by setting $y = B/A$, while the dotted curve is drawn by setting $y = B/A + 2c_{\text{dark}}/\overline{\langle L\rangle_{\mathcal{R}}} Ah\delta\nu$, where the temperature dependence of $B/A$ was fit by an exponential function as shown in Fig. S5, $\delta\nu$ is the frequency range corresponds to one line of vertically binned pixels in the CCD camera, $\overline{\langle L\rangle_{\mathcal{R}}} \equiv (1/N)\sum_{i=1}^{N}\langle L\rangle_{\mathcal{R}}^{(T_i)}$ represents the average of $\langle L\rangle_{\mathcal{R}}$ over temperatures $T_i = \{85, 90, \cdots, 300\text{ K}\}$,, and $c_{\text{dark}}$ is the counts due to the CCD's dark current whose average value is cancelled by the background measurement while it adds noise to the spectrum. The dotted curve explains the experimentally observed standard deviation $\sigma_A$ and the residual would be associated with the background counts from the surroundings of NDs such as the PDMS sheet. The noise due to $c_{\text{dark}}$ is non-negligible because the PL is spread over thousands of pixels in the CCD camera in the spectrometer.

### N. Derivation of the Function $g(y)$

Application of the theory given in Ref. [S2] to the case with squared-Lorentzian function gives

$$g(y) = \sqrt{\frac{f_2(y)}{f_1(y)f_2(y) - (f_3(y))^2}} \frac{\sqrt{\pi}\Gamma\left(\beta - \frac{1}{2}\right)}{\Gamma(\beta)} \quad (S15)$$

$$f_1(y) = \int_{-\infty}^{\infty} dx \left(\frac{(x^2+1)^\beta}{y(x^2+1)^\beta + 1}\right)\left(\frac{1}{(x^2+1)^\beta}\right)^2 \quad (S16)$$

$$f_2(y) = \int_{-\infty}^{\infty} dx \left(\frac{(x^2+1)^\beta}{y(x^2+1)^\beta + 1}\right)\left(\frac{x^2}{(x^2+1)^{\beta+1}}\right)^2 \quad (S17)$$

$$f_3(y) = \int_{-\infty}^{\infty} dx \left(\frac{(x^2+1)^\beta}{y(x^2+1)^\beta + 1}\right)\left(\frac{x^2}{(x^2+1)^{\beta+1}}\right), \quad (S18)$$

where $\beta = 2$ and $\Gamma(x)$ is the Gamma function. Instead of evaluating them analytically, we computed them numerically and fit the function $g(y)$ by a form $\sqrt{c_1 + c_2 y + c_3\sqrt{y^2 + y}}$ as shown in the Fig. S4, where $\{c_1, c_2, c_3\}$ are fitting parameters. The function was well fit by $[c_1, c_2, c_3] \simeq [2.00, 1.98, 0.763]$.

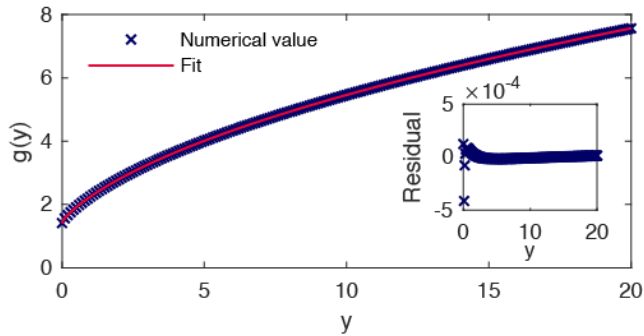

FIG. S4. Numerical evaluation of the function $g(y)$ and the fit. Inset shows the residuals.

### O. Calculation of the projected sensitivity

In our experiment, the ZPL counts rates were orders of magnitude smaller than those measured in the former study, where the ZPL counts rate from a single ND was observed to be $C_{\text{ZPL},1}^{(295\text{ K})} = 900$ kcps at $T = 295$ K [S1], in contrast to our measurement of $C_{\text{ZPL},2}^{(295\text{ K})} = 760$ kcps at $T = 295$ K. High sensitivity all-optical thermometry with $\eta = 300$ mK Hz$^{-1/2}$ was demonstrated with this high ZPL detection rate $C_{\text{ZPL},1}^{(295\text{ K})}$. To compare our result with the previous study, we define a projected sensitivity

$$\eta_{\text{proj}} = \sqrt{C_{\text{ZPL},2}^{(295\text{ K})}/C_{\text{ZPL},1}^{(295\text{ K})}}\,\eta \tag{S19}$$

and it is shown in the right axis of the Fig. 3(d) in the main text. Though the projected sensitivity only gives a rough estimate of a sensitivity given a higher detection efficiency of the PL, it shows our result is consistent with the previous report.

### P. Temperature Dependence of Sensitivity

From the equations (S12) and (S13), we get the rescaled sensitivity

$$\eta\sqrt{C_{\text{ZPL}}} \equiv \sigma_A\sqrt{C_{\text{ZPL}}\Delta t}\left|\frac{dT}{dA}\right| = \frac{g(y)}{2T\left(\frac{b}{a+bT^2}+\gamma\right)}. \tag{S19}$$

The rescaled sensitivity represents the minimum temperature difference that can be resolved by a single ZPL photon detection. We show the temperature dependence of the rescaled sensitivity in Fig. S5. Two lower bounds due to the models explained in the previous section are shown. The equation (S19) gives a low temperature behavior $\eta\sqrt{C_{\text{ZPL}}} \sim 1/T$, which is consistent with the experimental data shown in Fig. S5.

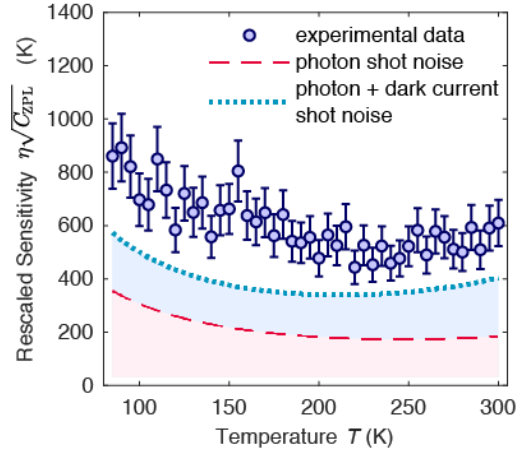

FIG. S5. Temperature dependence of the rescaled sensitivity $\eta\sqrt{C_{\text{ZPL}}}$. Two curves showing the lower bound due to the two limitations as explained in the Fig. 3 in the main text are shown. The rescaled sensitivity represents the minimum possible temperature difference that can be measured by a single ZPL photon detection.

Temperature dependence of the sensitivity can be derived from the equations (S4) and (S19), resulting in

$$\eta = \frac{g(y)}{2T\left(\frac{b}{a+bT^2}+\gamma\right)\sqrt{C_{\text{tot}}(\text{DWF})|_{T=0}}}\exp\left(\frac{1}{2}\gamma T^2\right). \tag{S20}$$

Under simplifying approximations $a \gg Tb^2$, $\gamma \gg b/a$ and $g(y) \simeq g(0) \simeq \sqrt{2}$, we get

$$\eta \simeq \frac{1}{T\gamma\sqrt{2C_{\text{tot}}(\text{DWF})|_{T=0}}}\exp\left(\frac{1}{2}\gamma T^2\right), \tag{S21}$$

which gives a minimum at $T = 1/\sqrt{\gamma}$.

### Q. Discussion of the Effect of the PDMS Sheet on the Sensitivity Measurement

Figure. 3(d) in the main text shows non-negligible effects of the PDMS sheet. This is mainly due to the temperature dependence of the absolute ZPL counts rate shown in the inset of Fig.3(c), which is largely affected by the optical transparency of the PDMS sheet that modifies the PL collection efficiency of our setup. Temperature dependence of the rescaled sensitivity shown in Fig. S5 supports this statement, since there are no observable dips/peaks in the figure. While the inset of Fig. 3(c) and Fig. 3(d) are affected by the existence of the PDMS sheet, the general tendency of these figures are expected to be due to the NV-center's intrinsic properties, since the temperature dependence of the total PL of NV centers below room temperatures are reported to be negligible [S9–S11], leading to the decrease of the ZPL counts rate with temperature increase, due to the Debye-Waller factor.

### R. Choice of the Base Temperature $T = 170$ K for the Temperature Imaging of YIG

We chose the base temperature of $T = 170$ K in the measurement of Fig. 4 in the main text. This is because there is a glass transition of PDMS at $T \simeq 150$ K. Below the glass transition of PDMS, the proximity of the NDs on the YIG surface is not ensured and the local temperature measurements become untrustworthy. Above the transition temperature, the NDs are in good contact with the YIG surface and they measure the local temperatures of the YIG. Note that the glass transition does not affect the main results in other parts of this report since the temperature gradient was not applied.

### S. Calibration of the temperature sensor

For the calibration of the temperature sensors, we conducted multiple scans of the spectrum at $T = 170$ K and $T = 180$ K by changing the base temperatures of the copper thermal sink. The average value and the variance of the fitting coefficients $\{A, B, \Theta, w, \nu_{ZPL}\}$ were extracted. Then we calculated the linear dependence to convert the value of $\{A, B, \Theta, w, \nu_{ZPL}\}$ into temperatures. The calibration was conducted for each spot in the array.

### T. Temperature Imaging of YIG

As mentioned in the main text, the temperature dependencies of the parameters $\{B, \Theta, w, \nu_{ZPL}\}$ in addition to $A$ were used for temperature sensing by taking the weighted average of the temperatures measured by fitting coefficients. In the Figure S6, the temperature dependencies of $\Theta$, $B/A$ and $\nu_{ZPL}$ are shown, where $B/A$ was fit by an exponential function which is empirical but the specific functional form does not matter in this report. The parameter $\Theta$ represents the slope of the exponential function in the fit of the phonon sideband. The value is different from the true temperature by a factor of order one, which is called Urbach's rule and similar dependencies are observed in many other materials [S12]. We note that the temperature dependence of this exponential tail can potentially be used as a temperature sensor below the liquid nitrogen temperatures for future applications.

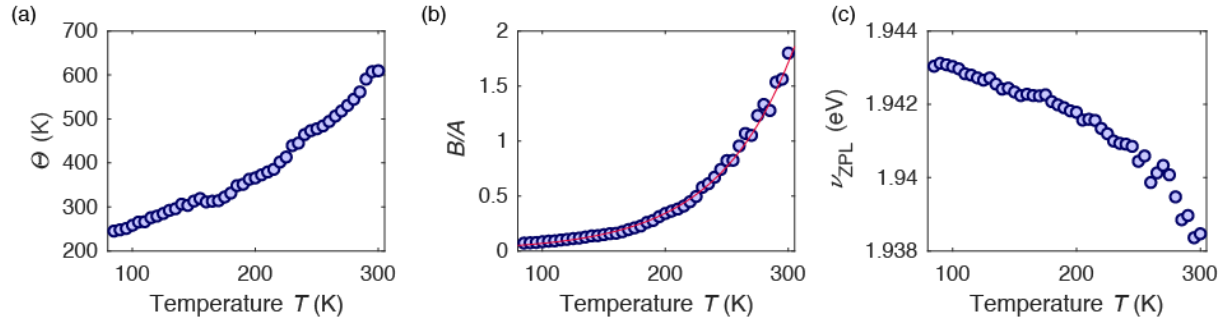

FIG. S6. Temperature dependencies of (a)$\Theta$, (b)$B/A$ and (c)$\nu_{ZPL}$. The ratio $B/A$ was fit by an exponential function in (b) with a solid (red) curve.

Before the temperature measurements, YIG was magnetized to one direction by applying a DC magnetic field. After taking temperature measurements on multiple spots in the ND array, the temperatures around the scanned spots are smoothly interpolated or extrapolated and shown in the Fig. 4(b) in the main text. The spots used in the temperature measurement is shown in Fig. S7, where the white circles represent the spots that were used.

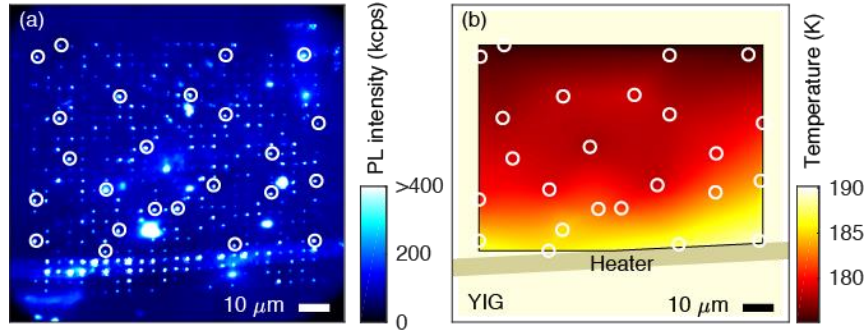

FIG. S7. Two-dimensional scan of PL and temperatures as shown in the Fig. 4 in the main text with circles representing the spots in the array that were used for the temperature measurement. Temperatures around the scanned spots were smoothly interpolated or extrapolated in (b).

### U. Discussion of the temperature profile

In the Fig. 4(c) in the main text, the YIG surface temperature $T(x)$ was fit by a logarithmic function
$$T(x) = -\xi \log(x - \zeta), \tag{S19}$$
where $x$ is the distance from the resistive heater and $\{\xi, \zeta\}$ are fitting parameters. A logarithmic function is used because the Green's function to the steady state two-dimensional diffusion equation with a single hear carrier is logarithmic. Since the wire has the length of 200 $\mu m$ and the center of the PDMS sheet was displaced to the left by approximately 45 $\mu$m due to experimental imperfection, there would be a deviation from the logarithmic function due to the imperfection of the two-dimensionality, i.e., the resistive heater is not infinitely long. A finite YIG thickness and the existence of an interface between YIG and GGG can also be a potential cause of the deviation from the logarithmic function. The deviation is, however, not clearly observed.

In addition, both phonons and magnons are the heat carriers in YIG. According to the coupled magnon-phonon heat transport theory [S13], the steady state phonon temperature profile $T_p(\mathbf{x})$ does not obey a simple Poisson equation, but it obeys
$$\left(\kappa_m \kappa_p \nabla^4 - g(\kappa_m + \kappa_p)\nabla^2\right)T_p = -(\kappa_m \nabla^2 - g)Q_p + g Q_m. \tag{S20}$$
Here we ignored, for simplicity, the spatial derivatives of the thermal conductivities, $\kappa_m$ and $\kappa_p$. Parameters $Q_p$ and $Q_m$ are the power densities of external heating absorbed by phonons and magnons, respectively [S14]. It is shown in the reference [S14], however, that the equation (S20) can be approximated to the Poisson equation in a regime where the phononic temperature gradient is dominant over the gradient of the magnon-phonon temperature difference. The observed logarithmic behavior of the phononic temperature profile supports this approximation and that the steady-state phononic temperature profile is not largely disturbed by magnons in YIG. For further study of the temperature profile of the YIG film, higher temperature sensitivity is required.